\documentclass[%
 reprint,
 twocolumn,
superscriptaddress,
nofootinbib,
 amsmath,amssymb,
 prd,
]{revtex4-2}

\usepackage{graphicx}
\usepackage{dcolumn}
\usepackage{bm}
\usepackage{color}
\usepackage{siunitx}
\usepackage{url,hyperref}
\usepackage[dvipsnames]{xcolor}
\usepackage{lineno}

\newcommand{\Ogw}{\Omega_\text{GW}}

\newcommand{\calP}{{\cal P}}

\newcommand{\pbar}{\bar{\cal P}}
\newcommand{\hf}[2]{\bar{H}_{#1}^{#2}(f; \{\vec{\theta^\prime}\})}
\newcommand{\hfinj}[2]{\bar{H}_{#1}^{#2}(f; \{\vec{\theta}^\mathrm{inj}\})}

\newcommand{\lmax}{\ell_\text{max}}

\newcommand{\orf}[2]{\gamma_{#1}^{#2}(f,t)}
\newcommand{\fisher}[2]{\Gamma_{#1}^{#2}}

\newcommand*{\diff}{\,\mathrm{d}}

\renewcommand{\eqref}[1]{Eq.~(\ref{#1})}
\newcommand{\eqsref}[2]{Eqs.~(\ref{#1}) and (\ref{#2})}
\newcommand{\figref}[1]{Fig.~\ref{#1}}
\newcommand{\figureref}[1]{Figure~\ref{#1}}
\newcommand{\secref}[1]{Sec.~\ref{#1}}
\newcommand{\appref}[1]{Appendix~\ref{#1}}

\usepackage[acronym]{glossaries}
\makeglossaries
\newacronym[plural=GWs,firstplural=gravitational waves (GWs)]{gw}{GW}{gravitational wave}
\newacronym[plural=BHs,firstplural=black holes (BHs)]{bh}{BH}{black hole}
\newacronym[plural=CBCs,firstplural=compact binary coalescences (CBCs)]{cbc}{CBC}{compact binary coalescence}
\newacronym[plural=MSPs,firstplural=millisecond pulsars (MSPs)]{msp}{MSP}{millisecond pulsar}
\newacronym[plural=SGWBs,firstplural=stochastic gravitational-wave background (SGWBs)]{sgwb}{SGWB}{stochastic gravitational-wave background}
\newacronym[plural=BFs,firstplural=Bayes factors (BFs)]{bf}{BF}{Bayes factor}
\newacronym{ligo}{LIGO}{Laser Interferometer Gravitational-wave Observatory}
\newacronym{lvk}{LVK}{LIGO Scientific, Virgo and KAGRA Collaboration}
\newacronym{o1}{O1}{the first observing run}
\newacronym{o2}{O2}{the second observing run}
\newacronym{o3}{O3}{the third observing run}
\newacronym{o3a}{O3a}{the first half of the third observing run}
\newacronym{o4}{O4}{the fourth observing run}
\newacronym{o5}{O5}{the fifth observing run}
\newacronym{lisa}{LISA}{Laser Interferometer Space Antenna}
\newacronym{snr}{SNR}{signal-to-noise ratio}
\newacronym{cmb}{CMB}{cosmic microwave background}
\newacronym{cgwb}{CGWB}{cosmic gravitational-wave background}
\newacronym{csd}{CSD}{cross spectral density}
\newacronym{psd}{PSD}{power spectral density}
\newacronym{bpl}{BPL}{broken power law}
\newacronym{pl}{PL}{power-law}
\newacronym{shd}{SHD}{spherical-harmonics decomposition}
\newacronym{orf}{ORF}{overlap reduction function}
\newacronym{gwtc}{GWTC}{Gravitational Wave Transient Catalog}
\newacronym{et}{ET}{Einstein Telescope}
\newacronym{ce}{CE}{Cosmic Explorer}

\begin{document}


\title{Bayesian parameter estimation\\for targeted anisotropic gravitational-wave background}

\author{Leo Tsukada}
\email{leo.tsukada@ligo.org}
\affiliation{Department of Physics, The Pennsylvania State University, University Park, Pennsylvania 16802, USA}
\affiliation{Institute for Gravitation and the Cosmos, The Pennsylvania State University, University Park, Pennsylvania 16802, USA}

\author{Santiago Jaraba}
\affiliation{Instituto de F\'isica Te\'orica UAM-CSIC, Universidad Aut\'onoma de Madrid, 28049 Madrid, Spain}

\author{Deepali Agarwal}
\affiliation{Inter-University Centre for Astronomy and Astrophysics (IUCAA), Pune 411007, India}

\author{Erik Floden}
\affiliation{School of Physics and Astronomy, University of Minnesota, Minneapolis, Minnesota 55455, USA}
\date{\today}

\begin{abstract}
  Extended sources of the stochastic gravitational backgrounds have been
  conventionally searched on the spherical-harmonics bases. The analysis during
  the previous observing runs by the ground-based gravitational-wave detectors,
  such as LIGO and Virgo, have yielded the constraints on the angular power
  spectrum $C_\ell$, yet it lacks the capability of estimating other parameters
  such as a spectral index.  In this paper, we introduce an alternative Bayesian
  formalism to search for such stochastic signals with a particular distribution
  of anisotropies on the sky.  This approach provides a Bayesian posterior of
  model parameters and also enables selection tests among different signal
  models.  While the conventional analysis fixes the highest angular scale
  \textit{a priori}, here we show a more systematic and quantitative way to
  determine the cutoff scale based on a Bayes factor, which depends on the
  amplitude and the angular scale of observed signals. Also, we analyze the
  third observing runs of LIGO and Virgo for the population of millisecond
  pulsars and obtain the 95 \% constraints of the signal amplitude, $\epsilon <
  2.7\times 10^{-8}$.
\end{abstract}

\maketitle


\section{\label{sec:intro} Introduction}
Since the dawn of \gls{gw} astronomy~\cite{gw150914}, a series of \gls{gwtc}
published by the \gls{lvk}~\cite{gwtc-1, gwtc-2,gwtc-2.1, gwtc-3} have confirmed
90 \gls{gw} signals in total from \glspl{cbc} in the Universe. In upcoming
observing runs by ground-based \gls{gw} detectors, the detector network will be
expanding with the Advanced \gls{ligo}~\cite{AdvancedLIGO} and
Virgo~\cite{AdvancedVirgo} as well as KAGRA~\cite{kagra} and
LIGO-India~\cite{ligo_india}. Additionally, the next generation of ground-based
detectors such as the \gls{et}~\cite{et1, et2, et3, et4} and the \gls{ce}~\cite{ce} have been
proposed to further increase the sensitivity to various types of \gls{gw}
sources.

A \gls{sgwb} is the incoherent superposition of \glspl{gw} emitted from many
sources that are too faint to be resolved individually (see
e.g.,~\cite{arianna_review} for the detailed review).  It is composed mainly of
astrophysical sources such as binary black holes and binary neutron stars \cite{tania,
Regimbau_2022,Banagiri:2020kqd,Payne:2020pmc,Stiskalek:2020wbj}, supernovae
\cite{2009MNRAS.398..293M,2010MNRAS.409L.132Z,PhysRevD.72.084001,PhysRevD.73.104024,marocmodi},
or ultralight bosons corotating around a \gls{bh}
\cite{Brito:2017wnc,Brito:2017zvb,Fan:2017cfw,Tsukada:2018mbp,palomba2019direct,sun2019search}.
Alternatively, cosmological sources can contribute to the \gls{sgwb}, which
include signals emitted during an inflationary
era~\cite{1994PhRvD..50.1157B,1979JETPL..30..682S,2007PhRvL..99v1301E,2012PhRvD..85b3525B,2012PhRvD..85b3534C,Lopez:2013mqa,1997PhRvD..55..435T,2006JCAP...04..010E,axion_inflation},
phase transitions in the early Universe~\cite{vonHarling:2019gme,Dev:2016feu,Marzola:2017jzl}, and primordial
\glspl{bh} \cite{MandicEA_2016,SasakiEA_2016,Wang:2016ana,Miller:2020kmv}.
These theoretical models, in general, predict a characteristic background
amplitude, spectral shape or angular distribution
\cite{contaldi_aniso,Jenkins:2018kxc,
Jenkins:2019uzp,Jenkins:2019nks,Bertacca:2019fnt,cusin_pitrou_analytic_expression_angular_power,Cusin:2017mjm,Cusin:2018rsq,Cusin:2019jpv,Pitrou:2019rjz,Canas-Herrera:2019npr,Geller:2018mwu}.
Therefore, one can not only detect but also, in principle, distinguish the
different models by comparing these signatures in observed signals.

In contrast with established detections of \glspl{cbc}, \glspl{sgwb} have not
yet been detected and hence it is one of the next milestones in the future
observing runs. Conventional searches for a \gls{sgwb} are mainly categorized
into two types: isotropic~\cite{s5_iso, o1_iso, o2_iso, o3_iso} and directional
searches~\cite{s5_directional, o1_directional, o2_directional, o3_directional,
asaf}.  While the former assumes isotropic energy distribution of \glspl{gw}
over the sky and estimates the overall amplitude of the \gls{sgwb}, the latter
targets for its anisotropic distribution. Regarding the directional searches,
two different methodologies have been adopted depending on signal models to
pursue, e.g., using radiometer
analyses~\cite{radio_method,Ballmer2006LIGOIO,Mitra_2008_radio_method2} for
pointlike sources or the \gls{shd} analysis~\cite{sph_methods} for extended
sources.

Inspired by the \gls{shd} analysis approach, our work presented in this paper
describes a Bayesian parameter estimation formalism targeted for an anisotropic
distribution of \gls{gw} energy. In the literature, similar targeted searches
for different anisotropic sources have been introduced~\cite{deepali_targeted,
kinematic_dipole}, both of which construct the maximum-likelihood estimator of
an overall amplitude of the \gls{gw} energy density and provide its upper limit.
Unlike these methods or the conventional \gls{shd} method which produces an
estimator for each spherical-harmonics mode of the energy distribution, our
formalism described here assumes an anisotropy model as known and infers other
model parameters, e.g., an energy spectrum or an overall amplitude, through
stochastic sampling. This can be seen as the extension of a parameter estimation
for an isotropic background~\cite{stochastic_pe_iso, pat_pe} by adding higher
spherical-harmonics modes in a signal model. As will be discussed in the rest of
the paper, this also allows us to perform a selection across different
anisotropy models and to optimize a spatial cutoff scale in the signal model.

This paper is structured as follows. First, \secref{sec:shd} provides an
overview of the conventional \gls{shd} analysis. Second, in
\secref{sec:formalism} we describe our parameter estimation formalism based on
the Bayesian framework. Following the procedure of data simulation and signal
injection described in \secref{sec:data}, from \secref{sec:stats} to
\secref{sec:lmax_opt} we show the results of various studies to assess the
parameter inference consistency, the model selection and the optimization of the
spatial cutoff scale. Subsequently, in \secref{sec:o3_zerolag} we analyze the real data
from \gls{o3} by the Advanced \gls{ligo} and Virgo based
on a population of \glspl{msp}. Finally, in \secref{sec:future} we discuss future
prospects on the precision of parameter estimation and the constraints on model
parameters by simulating several planned future detectors.

\section{\label{sec:shd} Anisotropic stochastic gravitational wave background}
  The anisotropy of the \gls{sgwb} can be expressed in terms of the
  dimensionless energy density
  $\Omega_{\mathrm{gw}}$~\cite{allen-romano,romano2017detection, arianna_review}
  \begin{align}
    \label{eq:omegagw1}
    \Omega_{\mathrm{gw}}(f, \hat{\Omega})\equiv\frac{f}{\rho_c}\frac{\diff^3\rho_\mathrm{gw}}{\diff f\diff^2\hat{\Omega}},
  \end{align}
  where $\diff^3\rho_\mathrm{gw}$ is the \gls{gw} energy per unit frequency $f$
  and solid angle $ \hat{\Omega} $, $ \rho_c $ is the critical energy density
  required to have a spatially flat universe and $ H_0 $ is the Hubble constant.
  Assuming that $ \Omega_{\mathrm{gw}} $ can be factorized into frequency and
  sky-direction dependent terms, i.e. $ H(f) $ and $ \mathcal{P}(\hat{\Omega}) $
  respectively, the above equation reads
  \begin{align}
    \label{eq:omegagw2}
    \Omega_{\mathrm{gw}}(f, \hat{\Omega})=\frac{2\pi^2}{3H_0^2}f^3H(f)\mathcal{P}(\hat{\Omega}).
  \end{align}
  This assumption has been shown~\cite{romano2017detection,allen-ottewill} to
  hold across the frequency range in which \gls{lvk} stochastic searches are
  most sensitive.  Most of the literature adopts a power-law form for the
  frequency spectrum $H(f)=(f/f_\mathrm{ref})^{\alpha-3}$, for example,  $
  \alpha=2/3 $ as predicted for a \gls{cbc}
  background~\cite{sesana2008stochastic}. This spectral model is used to perform
  a broadband search where one constructs sky maps by integrating different
  detector outputs over a broad range of frequencies.

  For the sky-position dependent term $ \mathcal{P}(\hat{\Omega}) $, we apply the \gls{shd},
  \begin{align}
  \mathcal{P}(\hat{\Omega})=\sum_{\ell,m} \mathcal{P}_{\ell m}Y_{\ell m}(\hat{\Omega}),
  \end{align}
  where $ Y_{\ell m}(\hat{\Omega}) $ is a spherical-harmonics function evaluated at
  the sky position $ \hat{\Omega} $. One uses the \gls{shd} to search for extended
  sources with a large angular scale. We first construct the \textit{dirty map} $ X_\nu
  $, which is essentially a cross-correlation between different detector outputs, and its
  covariance matrix $ \fisher{\mu\nu}{} $ projected onto spherical-harmonics
  bases.  We then produce a \textit{clean map} as a nonbiased estimator of $
  \mathcal{P}_{\ell m} $, deconvolving the dirty map by the covariance matrix
  \begin{align}
    \label{eq:clean_map}
    \hat{\mathcal{P}}_{\mu}=\sum_{\nu}(\fisher{}{-1})_{\mu\nu}X_{\nu}.
  \end{align}
  Here $ \mu,\nu $ subscripts represent a spherical-harmonics mode, i.e. $\mu
  \equiv ( \ell,m ) $, and $ \Gamma^{-1} $ is the inverted covariance matrix.

  To constrain anisotropies of a \gls{sgwb}, we introduce the following estimator of the
  angular power spectrum
  \begin{align}
  \hat{C_\ell}=\left(\frac{2\pi^2f^3_\mathrm{ref}}{3H_0^2}\right)^2\frac{1}{2\ell+1}\sum_{m}\left[|\hat{\mathcal{P}}_{\ell m}|^2-(\Gamma^{-1})_{\ell m, \ell^\prime m^\prime}\right],
  \end{align}
  which we compare to theoretical predictions.

\section{\label{sec:formalism} Formalism}
  \subsection{\label{sec:formalism-bayes} Bayesian inference}
    Bayesian inference has been employed to estimate model parameters in the
    context of \gls{gw} astronomy, such as signals from \glspl{cbc}~\cite{lalinference,
    bilby1, bilby2} and an isotropic \gls{gw} background~\cite{stochastic_pe_iso,
    pat_pe}. The \gls{csd} at the frequency $f$ and the timestamp $t$ between
    outputs of the two detectors is defined as
    \begin{align}
      C(f, t)\equiv \frac{2}{\tau}\tilde{s}_1(f, t)\tilde{s}_2^*(f, t)
    \end{align}
    where $\tilde{s}(f, t)$ is a short-term Fourier transform of the time series
    $s(t)$ within an interval $[t-\tau/2, t+\tau/2]$. Let the anisotropic
    background depend on a set of model parameters $\{\vec{\theta}\}$. In the
    presence of the anisotropic background, the \gls{csd} on a two-dimensional
    $(f, t)$ pixel map has the mean~\cite{sph_methods}
    \begin{align}
      \label{eq:csd_mean}
      \langle C(f, t) \rangle = \orf{\mu}{}\mathcal{P}_{\mu}(f, \{\vec{\theta}\}),
    \end{align}
    where $\orf{\mu}{}$ denotes the \gls{orf}~\cite{ORF} projected onto the
    spherical-harmonics basis represented by $\mu=(\ell,
    m)$~\cite{allen-ottewill}, and $\mathcal{P}_{\mu}(f, \{\vec{\theta}\})$ is a
    generic form of a spectral model with the anisotropic distribution.  The Greek subscript implies the
    summation across the spherical-harmonics modes. The covariance of this
    \gls{csd} in the weak-signal approximation reads
    \begin{align}
      \label{eq:csd_cov}
      \langle |C(f, t)|^2 \rangle - |\langle C(f, t) \rangle |^2\approx \frac{P_1(f)P_2(f)}{\tau \Delta f},
    \end{align}
    where $P_i(f)$ is the \gls{psd} of the $i$th detector and $\Delta f$ is the
    frequency resolution.  Hence, these properties fully specify the Gaussian
    distribution, which represents the probability of obtaining the observable
    for a given signal model $\mathcal{M}$ and its relevant parameters,
    $p(\left\{C_{ft}\right\} \left|\{\vec{\theta}\} ; \mathcal{M}\right.)$.  In
    the Bayesian context, this is equivalent to the \textit{likelihood} and one
    aims to estimate the model parameters based on the \textit{posterior}
    distribution
    \begin{align}
      \label{eq:bayes_theorem}
        p\left(\left.\vec{\{\theta\}} \right|\{C_{f,t}\}; \mathcal{M}\right) = \frac{p\left(\left\{C_{f t}\right\} \left|\{\vec{\theta}\} ; \mathcal{M}\right.\right)p(\vec{\{\theta\}})}{p(\left\{C_{f t}\right\}|\mathcal{M})},
    \end{align}
    where $p(\vec{\{\theta\}})$ is a \textit{prior} distribution of the model
    parameters.  Note that $p(\left\{C_{f t}\right\};\mathcal{M})$ in
    the denominator is called the Bayesian \textit{evidence}
    \begin{align}
      \label{eq:evidence}
      p(\left\{C_{f t}\right\}|\mathcal{M}) = \int \boldsymbol{\diff}\vec{\{\theta\}}p\left(\left\{C_{f t}\right\} \left|\{\vec{\theta}\} ; \mathcal{M}\right.\right)p(\vec{\{\theta\}}),
    \end{align}
    which is used for a selection test across different hypotheses [see \secref{sec:model_selection}].

    \subsection{\label{sec:formalism-plm} $\mathcal{P}_{\ell m}$-specific Bayesian analysis}
    We further make several assumptions to simplify \eqref{eq:csd_mean};
    First, we decouple the frequency dependence from the anisotropy of the
    signal model.  Second, the only free parameter relevant to the anisotropy is
    its overall amplitude $\epsilon$. In other words, the signal model can be
    factorized as follows
    \begin{align}
      \mathcal{P}_{\ell m}(f, \{\vec{\theta}\}) = \epsilon \hf{}{} \pbar_{\ell m},
    \end{align}
    where $\hf{}{}$ and $\pbar_{\ell m}$ are normalized such that
    \begin{align}
      \label{eq:H_P_norm}
      \hf{}{}\equiv\frac{H(f; \{\vec{\theta^\prime}\})}{H(f_\mathrm{ref}; \{\vec{\theta^\prime}\})},\qquad \pbar_{\ell m} \equiv \frac{\mathcal{P}_{\ell m}}{\mathcal{P}_{00}}
    \end{align}
    respectively. Although one can relax these assumptions and consider a signal
    model with a more generic form, $\mathcal{P}_{\ell m}(f, \{\vec{\theta}\})$,
    in this paper we strict ourselves to this simplified case to perform
    $\mathcal{P}_{\ell m}$-specific Bayesian analysis so that $\mathcal{M}\rightarrow\pbar_{\ell m}$, and leave its
    generalization as future work. Lastly, we set the cutoff on spatial scale
    characterized by $\lmax$ and hence the subscript in \eqref{eq:csd_mean}
    implies
    \begin{align}
      \orf{\mu}{}\mathcal{P}_\mu = \sum_{\ell=0}^{\lmax}\sum^{\ell}_{m=-\ell}\orf{\ell m}{}\mathcal{P}_{\ell m}.
    \end{align}

    With these assumptions, the likelihood obeys the Gaussian distribution as below
    \begin{widetext}
      \begin{align}
        \label{eq:likelihood}
        p\left(\left\{C_{ft}\right\} \left|\{\epsilon, \vec{\theta^\prime}\} ;\lmax, \pbar_{\ell m}\right.\right)
          \propto\exp\left\{-\frac{1}{2}\sum_{f, t}\frac{\left|C(f, t)-\epsilon \hf{}{} \orf{\mu}{} \pbar_\mu\right|^{2}}{P_{1}(f, t) P_{2}(f, t)}
          \right\}.
      \end{align}
    \end{widetext}
    Note that in the case of $\lmax=0$, this likelihood expression reproduces
    the one for the isotropic stochastic background
    analysis~\cite{stochastic_pe_iso, pat_pe}. Also for the rest of this paper,
    the amplitude factor $\epsilon$ is normalized by
    $(\frac{2\pi}{3H_0^2}f_\mathrm{ref}^3\sqrt{4{\pi}})^{-1}$ so that this
    factor corresponds to $\hat{\Omega}_0$ in the isotropic search~\cite{s5_iso,
    o1_iso, o2_iso, o3_iso}. Therefore, this formalism serves as a natural
    extension of parameter inference performed for the isotropic background. 
    
    While this formalism yields results dependent on a specific anisotropy
    modeled by $\pbar_{\ell m}$, one can benefit from novel features as
    summarized below. First, the Bayesian evidence shown in \eqref{eq:evidence}
    allows one to perform a model selection test by comparing those between
    different $\hf{}{}$ or $\pbar_{\ell m}$ models.  Second, the anisotropy model
    $\pbar_{\ell m}$ also depends on the angular scale cutoff one imposes, and
    this can be optimized similarly by evaluating the Bayesian evidence across
    different $\lmax$ values [see \secref{sec:lmax_opt}]. Third, since
    \eqref{eq:likelihood} essentially computes the residual in the CSD $C(f,
    t)$, the likelihood shown in \eqref{eq:likelihood} does not involve the
    inversion of a covariance matrix $\fisher{\mu\nu}{}$.  In other words, this
    formalism is free from the regularization of an ill-conditioned covariance
    matrix, which otherwise would suffer from extra complexity just like the
    conventional \gls{shd} analysis~\cite{sph_methods}.

    For a given dataset and a signal model to recover, the pipeline
    stochastically samples over a multidimensional parameter space and
    iteratively evaluates the likelihood based on \eqref{eq:likelihood} for each
    sample until the posterior distribution can be sufficiently constructed.
    Regarding this implementation, we adopt a nested sampling algorithm
    \texttt{Dynesty}~\cite{dynesty}, implemented in the \texttt{Bilby}
    package.~\cite{bilby1, bilby2}

\section{\label{sec:data} Data simulation}
    The studies we conduct in the subsequent sections simulate a large set of
    noise data, e.g., the \gls{csd} $C(f, t)$, to assess statistical
    results. Here, we summarize our procedure of this data simulation and
    describe how it will be used for likelihood evaluation.
    \subsection{\label{sec:data-noise} Simulating noise data products} Given a
    set of the model parameters $\{\epsilon, \vec{\theta}\}$ and the signal
    model to recover, $\{\hf{}{}, \pbar_\mu\}$, the full expression of
    \eqref{eq:likelihood} can be written as
    \begin{widetext}
      \begin{align}
        \label{eq:likelihood2}
        2\ln\left[p\left(\left\{C_{ft}\right\} \left|\{\epsilon,
        \vec{\theta^\prime}\} ;\lmax, \pbar_{\ell m}\right.\right)\right]
        =-\ln\left[(2\pi)^d|\Sigma_{CC^\prime}|\right]
        - \sum_{f, t}\left\{\frac{\tau\Delta f|C(f, t)|^{2}}{P_{1}(f, t) P_{2}(f, t)}\right\}
        + 2\epsilon\mathrm{Re}[\pbar_{\mu}^{*} X_{\mu}]
        - \epsilon^{2}\pbar_{\mu}^{*} \fisher{\mu\nu}{} \pbar_{\nu},
      \end{align}
    \end{widetext}
    where $X_\mu,\fisher{\mu\nu}{}$ are defined as
    \begin{align}
      \label{eq:dirty_map}
        X_{\mu} &=\sum_{f} \sum_{t} \orf{\mu}{*} \frac{\tau\Delta f\hf{}{}}{P_{1}(f, t) P_{2}(f, t)} C(f, t)
    \end{align}
    and
    \begin{align}
      \label{eq:fisher_matrix}
        \fisher{\mu\nu}{} &=\sum_{f} \sum_{t} \orf{\mu}{*} \frac{\tau\Delta f\hf{}{2}}{P_{1}(f, t) P_{2}(f, t)} \orf{\nu}{},
    \end{align}
    respectively. In the weak signal limit, the covariance matrix of the
    \gls{csd}, $\Sigma_{CC^\prime}$, can be approximated as a diagonal matrix
    whose elements are given by \eqref{eq:csd_cov} with its dimension being
    \begin{align}
      d = (\mathrm{\#\ of\ frequency\ bins})\times(\mathrm{\#\ of\ time\ segments}).
    \end{align} The likelihood in \eqref{eq:likelihood2} is evaluated for a
    single baseline formed by a detector pair. For a general configuration of
    multiple baselines represented by $\boldsymbol{D}$, assuming uncorrelated noise
    among the detectors, the joint likelihood is given by a product of
    individual likelihood values computed for each distinct detector pair $\{IJ\}$
    \begin{align}
      \begin{split}
    p&\left(\left\{C_{f, t}^{\boldsymbol{D}}\right\} \left|\{\epsilon,
    \vec{\theta^\prime}\} ;\lmax, \pbar_{\ell m}\right.\right) =\\
    &\qquad\qquad\prod_I\prod_{I>J}p\left(\left\{C_{f, t}^{IJ}\right\} \left|\{\epsilon,
    \vec{\theta^\prime}\} ;\lmax, \pbar_{\ell m}\right.\right)
      \end{split}
    \end{align}

    Since the dirty map and the Fisher matrix need to be constructed based on
    \eqsref{eq:dirty_map}{eq:fisher_matrix} every time a
    pipeline draws a parameter sample during its stochastic sampling process,
    the two-dimensional integration over frequencies and time segments can be a
    computational bottleneck. Hence, after synthesizing $C(f,t)$ based on the
    colored Gaussian noise, we precompute and store the time integration part
    in \eqsref{eq:dirty_map}{eq:fisher_matrix}, i.e.
    \begin{align}
      \label{eq:dirty_map2}
        X_{\mu} &=\sum_{f}\hf{}{}\underbrace{\sum_{t} \frac{\tau\Delta f\orf{\mu}{*} C(f, t)}{P_{1}(f, t) P_{2}(f, t)}}_\mathrm{precomputed}\\
      \label{eq:fisher_matrix2}
        \fisher{\mu\nu}{} &=\sum_{f} \hf{}{2}\underbrace{\sum_{t} \frac{\tau\Delta f\orf{\mu}{*} \orf{\nu}{}}{P_{1}(f, t) P_{2}(f, t)}}_\mathrm{precomputed} ,
    \end{align}
    including other static data products such as $|\Sigma_{CC^\prime}|$ and
    \begin{align}
      \label{eq:csdsq_whiten}
      \sum_{f, t}\left\{\frac{\tau\Delta f|C(f, t)|^{2}}{P_{1}(f, t) P_{2}(f, t)}\right\}.
    \end{align}
  \subsection{\label{sec:data-injection} Signal injection}
  Following the pre-computation approach described above, we discuss how
  \eqref{eq:likelihood2} should be modified in the presence of a background
  signal. Specifically, we consider the background signal which contributes to the
  \gls{csd} such that
  \begin{align}
    \label{eq:csd_inj}
    C(f, t)= C_n(f, t) + \epsilon^\mathrm{inj}\hfinj{}{\mathrm{inj}}\orf{\mu}{}\pbar^\mathrm{inj}_{\mu},
  \end{align}
  where $C_n(f, t)$ is the noise \gls{csd} with zero mean. After
  substituting \eqref{eq:csd_inj} into \eqref{eq:likelihood2} with some
  expansion, the following terms will appear in the likelihood
  \begin{widetext}
  \begin{align}
    \label{eq:csdsq_whiten_inj}
    - 2\epsilon^\mathrm{inj}\mathrm{Re}\left[(\pbar^\mathrm{inj}_{\mu})^*X_\mu^\mathrm{inj}\right]
    - \left(\epsilon^\mathrm{inj}\right)^2(\pbar^\mathrm{inj}_\mu)^*\fisher{\mu\nu}{\mathrm{inj}}\pbar^\mathrm{inj}_\nu
    + 2\epsilon\epsilon^\mathrm{inj}\mathrm{Re}\left[\pbar^*_\mu\fisher{\mu\nu}{(c)}\pbar_\nu^\mathrm{inj}\right],
  \end{align}
  \end{widetext}
  where $X^\mathrm{inj}_{\mu}$ and $\fisher{\mu\nu}{\mathrm{inj}}$ follow the
  same definition as \eqsref{eq:dirty_map}{eq:fisher_matrix} except replacing
  $\hf{}{}$ and $C(f, t)$ with $\hfinj{}{\mathrm{inj}}$ and $C_n(f, t)$,
  respectively. We note that the last term in \eqref{eq:csdsq_whiten_inj}
  involves the \textit{coupled} Fisher matrix, which reads
    \begin{align}
      \label{eq:fisher_matrix_couple}
        \fisher{\mu\nu}{(c)} &=\sum_{f} \sum_{t} \orf{\mu}{*} \frac{\tau\Delta f\bar{H}\bar{H}^\mathrm{inj}}{P_{1}(f, t) P_{2}(f, t)} \orf{\nu}{}.
    \end{align}
  As we will describe in the subsequent section, the validity of this injection
  scheme can be shown by the statistical consistency in the injection recovery,
  e.g.,  \figref{fig:pp_plot}.

\section{\label{sec:stats} Statistical studies}
In order to assess the statistical consistency of the injection recoveries, we
perform injection campaigns using synthesized background signals with a specific
anisotropy model, $\bar{\calP}_{\ell m}$. To begin with, we summarize the setup
of the injection set, simulated data, and the parameter inference. Finally, the
implication from the injection campaigns will be described in
\secref{sec:stats-pp_plot}.
\subsection{\label{sec:stats-setup} Setup}
  \begin{itemize}
    \item Dataset:

    Following the procedure in \secref{sec:data}, we simulate unfolded dataset for
    a one-year observation divided into 192-second segments with the frequency
    resolution of \SI{0.25}{\hertz} starting from \SI{20}{\hertz} to \SI{500}{\hertz}. The \gls{csd}
    is produced from the cross-correlation between the two LIGO detectors with
    the projected O4 sensitivity ($\sim$\SI{190}{Mpc} of the binary inspiral
    range) shown in~\cite{observingScene, noise_curve}.
    \item $\pbar_{\ell m}$ model:

    As a toy model, we set $\pbar_{\ell m}$ to be a mock Galactic plane shown in
    \figref{fig:galactic_plane}. When injecting and recovering this model, we
    adopt $\lmax$ value consistent between the injection and recovery, ranging across
    3, 5 and 7. We choose the highest $\lmax=7$ because we find that the spherical-harmonics
    components above it only have negligible contribution to the total
    anisotropies. Also, note that the injected anisotropies with these $\lmax$ values
    appear to be blurred compared to \figref{fig:galactic_plane}.
    \begin{figure}[h]
      \centering
      \includegraphics[width=.9\columnwidth]{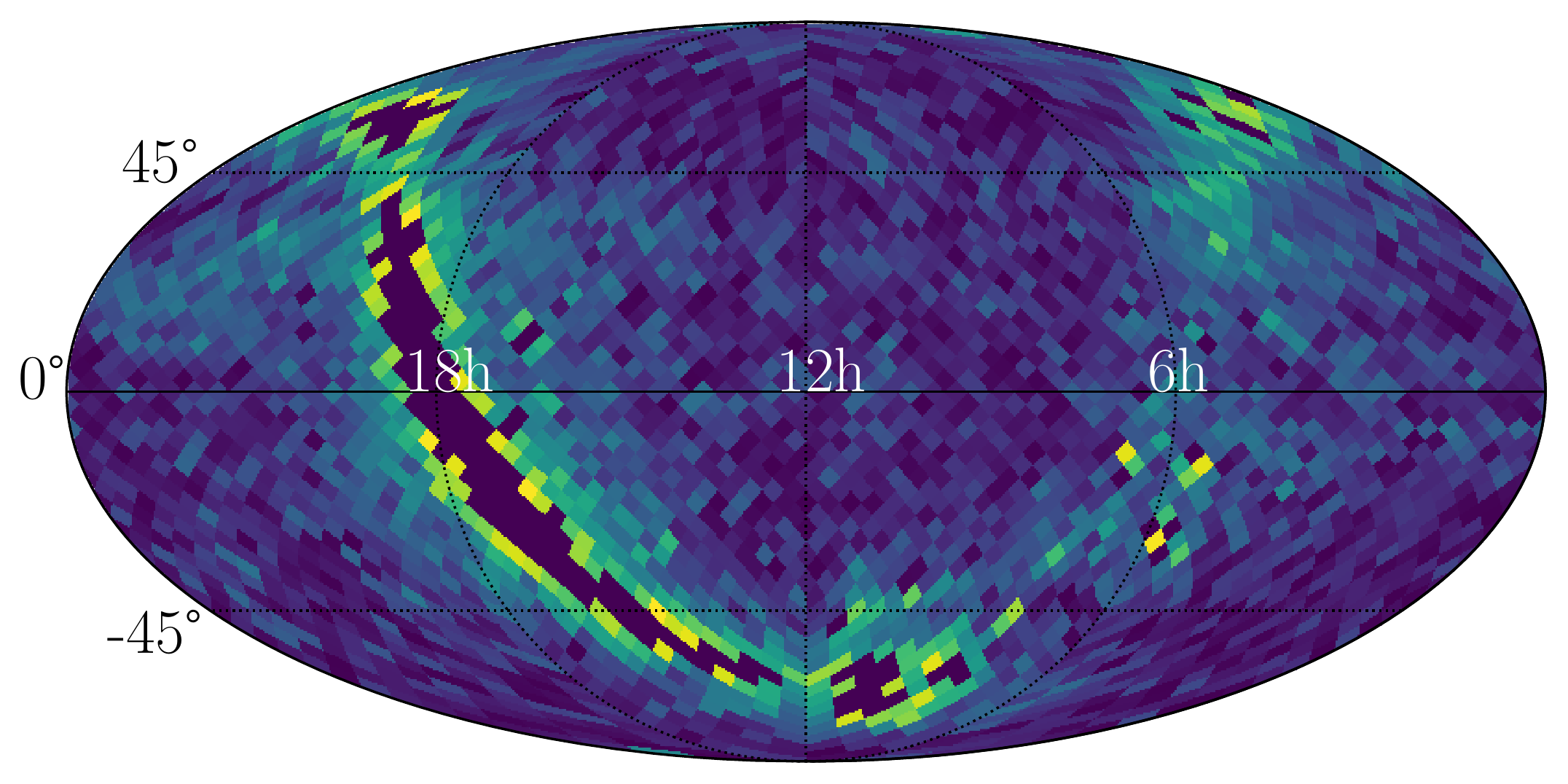}
      \caption{Distribution of a mock Galactic plane visualized by Mollweide
      projection with \texttt{HEALPix}~\cite{HEALPix} pixelization of
      $n_\mathrm{side}=16$. The brighter color represents larger energy
      density.}
      \label{fig:galactic_plane}
    \end{figure}
    \item $\bar{H}(f)$ model:

    We assume $\bar{H}(f)$ is a simple \gls{pl} model, i.e.
    \begin{align} \label{eq:H_PL} H_{\rm PL}(f;\alpha)=\left(\frac{f}{f_{\rm
    ref}}\right)^{\alpha-3} \end{align} where $f_\mathrm{ref}=\SI{25}{\hertz}$
    and the ``$-3$'' in the exponent is added so that $\Ogw\propto f^\alpha$ [see
    \eqref{eq:omegagw2}].

    \item Injections and prior distributions:

    Given the $\bar{H}(f)$ model above, the signal model has the two free
    parameters to infer, $(\epsilon, \alpha)$. In order to obtain a reasonable
    P-P plot, we adopt the same distributions consistently for both random draws
    of the injected parameters and the prior distributions of the same set of
    parameters for recovery as follows; for $\epsilon$, a log-uniform
    distribution in $10^{-8}$ to $10^{-5}$, and for $\alpha$, a Gaussian
    distribution with the mean of 5 and the standard deviation of 0.5. While the
    result of individual injection recoveries is found to be robust against
    changes in prior distributions, we select the above distributions so that
    most of the injections can be detected with great significance and their
    injected parameters are precisely inferred.
  \end{itemize}
\subsection{\label{sec:posterior_ex} Injection recovery}
\label{sec:stats-recovery}
\begin{figure}[h]
    \centering
    \includegraphics[width=\columnwidth]{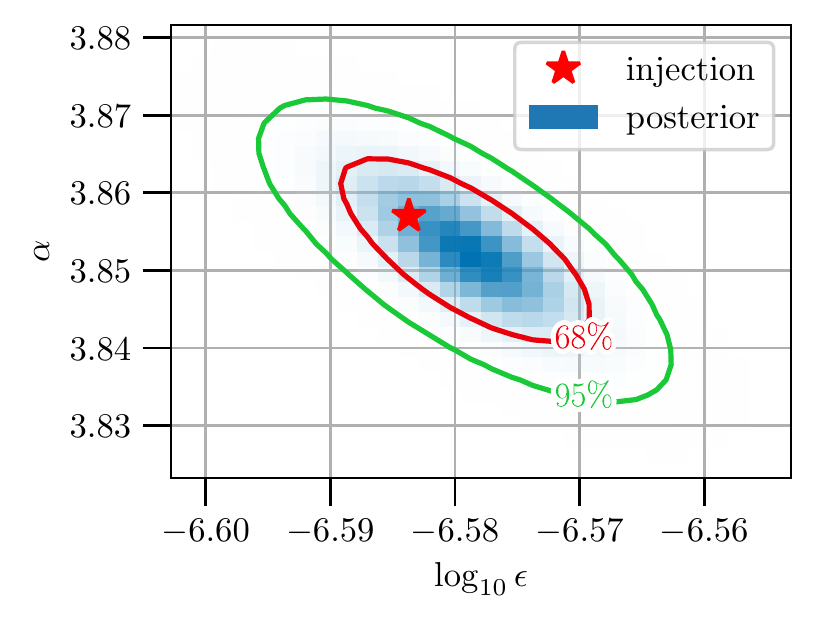}
    \caption{Posterior distribution for one of the detected injections. The
    injected values are $(\log_{10}\epsilon, \alpha)=(-6.584,3.857)$ and the
    \gls{bf} is $\sim24000$.}
    \label{fig:ex_post_detected}
  \end{figure}

  We follow the injection scheme described in \secref{sec:data-injection} and
  attempt to recover the injected values of the parameters by constructing the
  posterior distribution in the multidimensional parameter space. Here, we
  demonstrate two examples of injection recoveries performed with $\lmax=7$. In
  one case, an injection is loud enough to be detected and the parameters are
  precisely inferred. \figureref{fig:ex_post_detected} shows that the both
  parameters of $(\log_{10}\epsilon, \alpha)$ are inferred to the precision of
  $0.01$ and the injected values (the red star) are located close to the bulk of
  the posterior samples, staying inside the 68\% level of contours. As will be
  discussed later, the slight deviation of the injected values from the peak of
  the posterior distribution can be explained by the statistical error.

\begin{figure}[h]
    \centering
    \includegraphics[width=\columnwidth]{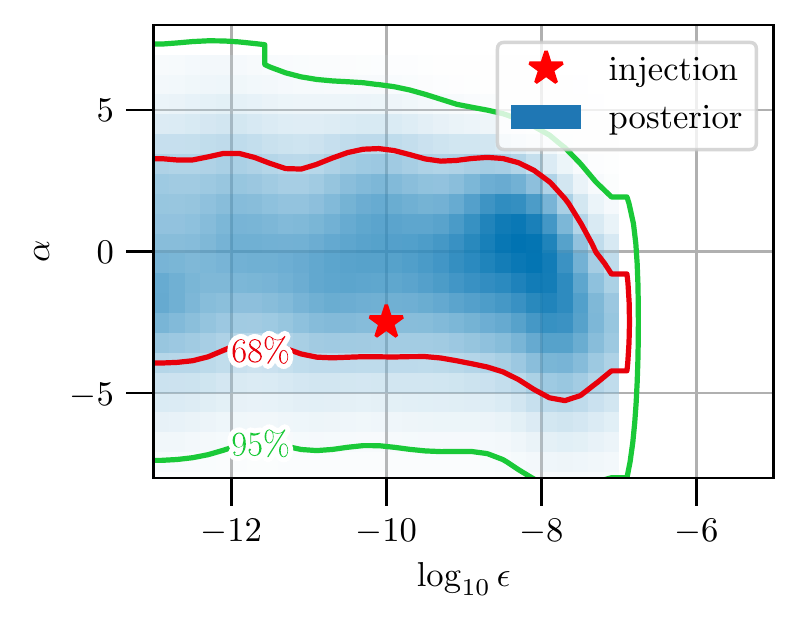}
    \caption{Posterior distribution for one of the nondetected injections. The
    injected values are $(\log_{10}\epsilon, \alpha)=(-10.0,-2.5)$ and the
    \gls{bf} is $\sim-0.35$.}
    \label{fig:ex_post_nondetected}
  \end{figure}

  In contrast, the other case makes an injection with a smaller amplitude such
  that the injection is not recovered with sufficient significance and that
  instead the posterior distribution provides the constraints on the inferred
  parameters. \figureref{fig:ex_post_nondetected} indicates the excluded region of the
  two parameters outside the contours. We note that, in order to thoroughly
  cover the parameter space, the prior distribution for this particular
  injection recovery is chosen to be a log-uniform distribution from $10^{-13}$ to
  $10^{-5}$ for $\epsilon$, and a Gaussian distribution with the mean of 0 and
  the standard deviation of 3.5 for $\alpha$. The analysis with this
  configuration is performed separately from the setup described in
  \secref{sec:stats-setup} for the purpose of demonstrating the parameter
  constraints.
\subsection{\label{sec:stats-pp_plot} Probability-probability plot}
\begin{figure}[h]
    \centering
    \includegraphics[width=\columnwidth]{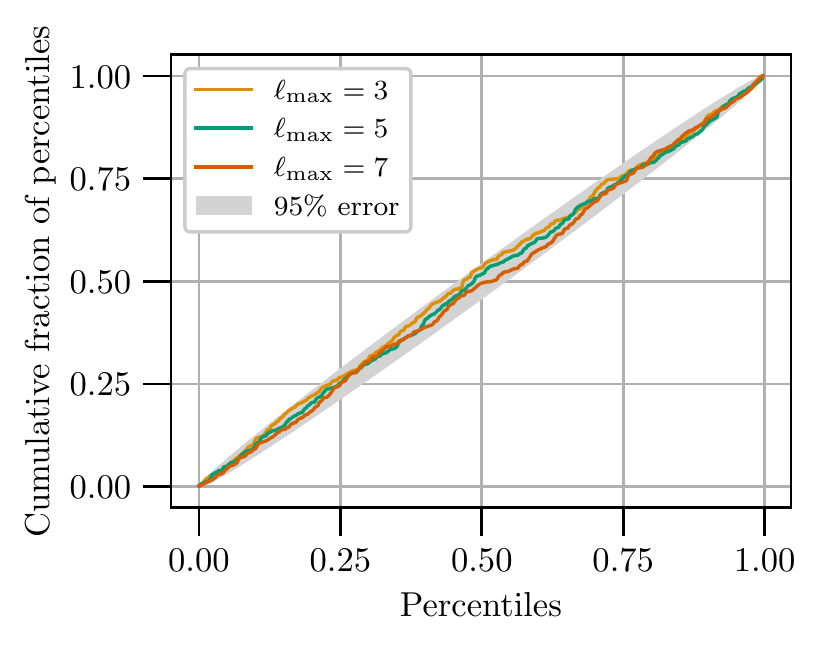}
    \caption{Probability-probability plot for 500 of the injection recoveries.
    Different colors indicate each value of $\lmax=3,5,7$. The gray region is
    the 95\% confidence region expected from the Poisson fluctuation.}
    \label{fig:pp_plot}
  \end{figure}

We perform 500 injection recoveries using $\pbar_{\ell m},\bar{H}(f)$ models and
the prior distributions described in \secref{sec:stats-setup}. Given the
posterior samples of each recovery, we compute a percentile of the injected
parameters with regard to the posterior probability. For example, if the
injected values in \figref{fig:ex_post_detected} sit on the red contour, we
would assign 68th percentile for the injection recovery. A collection of these
percentiles in turn provides the cumulative fraction at each percentile value in
the ascending order among the 500 injections. Eventually, we obtain the recovery
database, each of which contains the percentile and the cumulative density. We
plot them in \figref{fig:pp_plot}, which is referred to as a
probability-probability~(P-P) plot. Since the posterior distributions without
systematic error yield percentiles uniformly distributed between 0 and 1, a
trace in the plot is expected to stay close to the diagonal with some
statistical fluctuation. \figureref{fig:pp_plot} shows that the traces for all
the three cases of $\lmax=3,5,7$ stay within the 95\% credible error region.
This demonstrates the validity of data simulation, injection scheme and
posterior construction. Throughout this study, we consistently use the same
$\lmax$ value for both the injection and its recovery. However, the $\lmax$
value for the recovery can be different from that for the injection and we
discuss the potential bias caused by the $\lmax$ inconsistency in
\appref{app:iso_bias}.

\section{\label{sec:model_selection} Model selection}
  It is sometimes challenging to correctly interpret the underlying model
  from observed data. For instance, a \gls{bpl} $\hf{}{}$ model with similar
  exponents could easily be mistaken by a simple \gls{pl} model, especially for
  low \gls{snr}. As mentioned in \secref{sec:formalism-bayes}, this formalism
  allows us to quantitatively compare the preference across multiple signal
  models. Here, we will demonstrate this aspect of the pipeline using toy
  models.

  \subsection{\label{sec:model_selection-bf} Bayes factor}
    The fundamental tool to distinguish between two models within the Bayesian
    framework is the odds ratio. Given two models $\mathcal{M}_1$ and
    $\mathcal{M}_2$ and the data $D$, it is defined as
    \begin{equation}
      \mathcal{O}^{\mathcal{M}_1}_{\mathcal{M}_2}=\frac{p(\mathcal{M}_1|D)}{p(\mathcal{M}_2|D)}=\frac{p(D|\mathcal{M}_1)}{p(D|\mathcal{M}_2)}\frac{\pi(\mathcal{M}_1)}{\pi(\mathcal{M}_2)},
    \end{equation}
    where $\pi(\mathcal{M}_i)$ is the prior probability for model
    $\mathcal{M}_i$, $i=1,2$. Since, in our case, the prior odds for the both
    models are assumed to be equal, the odds ratio reduces to the \gls{bf},
    \begin{equation}
      \mathcal{B}^{\mathcal{M}_1}_{\mathcal{M}_2}=\frac{p(D|\mathcal{M}_1)}{p(D|\mathcal{M}_2)}.
    \end{equation}
    We assess the statistical significance of a model relative to the
    other in terms of the \gls{bf}. The evidence in favor of the model
    $\mathcal{M}_1$ is recognized to be strong around
    $\mathcal{B}^{\mathcal{M}_1}_{\mathcal{M}_2}\sim 10$ and decisive from $\sim
    100$~\cite{Kass1995}.

    Our tests proceed as follows: First, we synthesize the noise \gls{csd} from
    the cross-correlation between the two \gls{ligo} detectors with the
    projected O4 sensitivity. Then, we inject a signal simulated from a certain
    model $\mathcal{M}_1$ into the synthetic dataset. We recover it both with
    $\mathcal{M}_1$ and another model $\mathcal{M}_2$ involving different
    $\bar{H}(f)$ or $\pbar_{\ell m}$ model.  Finally, we compute the \gls{bf}
    for different sets of injected parameters until we obtain a heatmap showing
    a distribution of the \gls{bf} over the parameter space.

  \subsection{\label{sec:model_selection-hf} $\hf{}{}$ model selection}
\begin{figure}[h]
      \centering
      \includegraphics[width=\columnwidth]{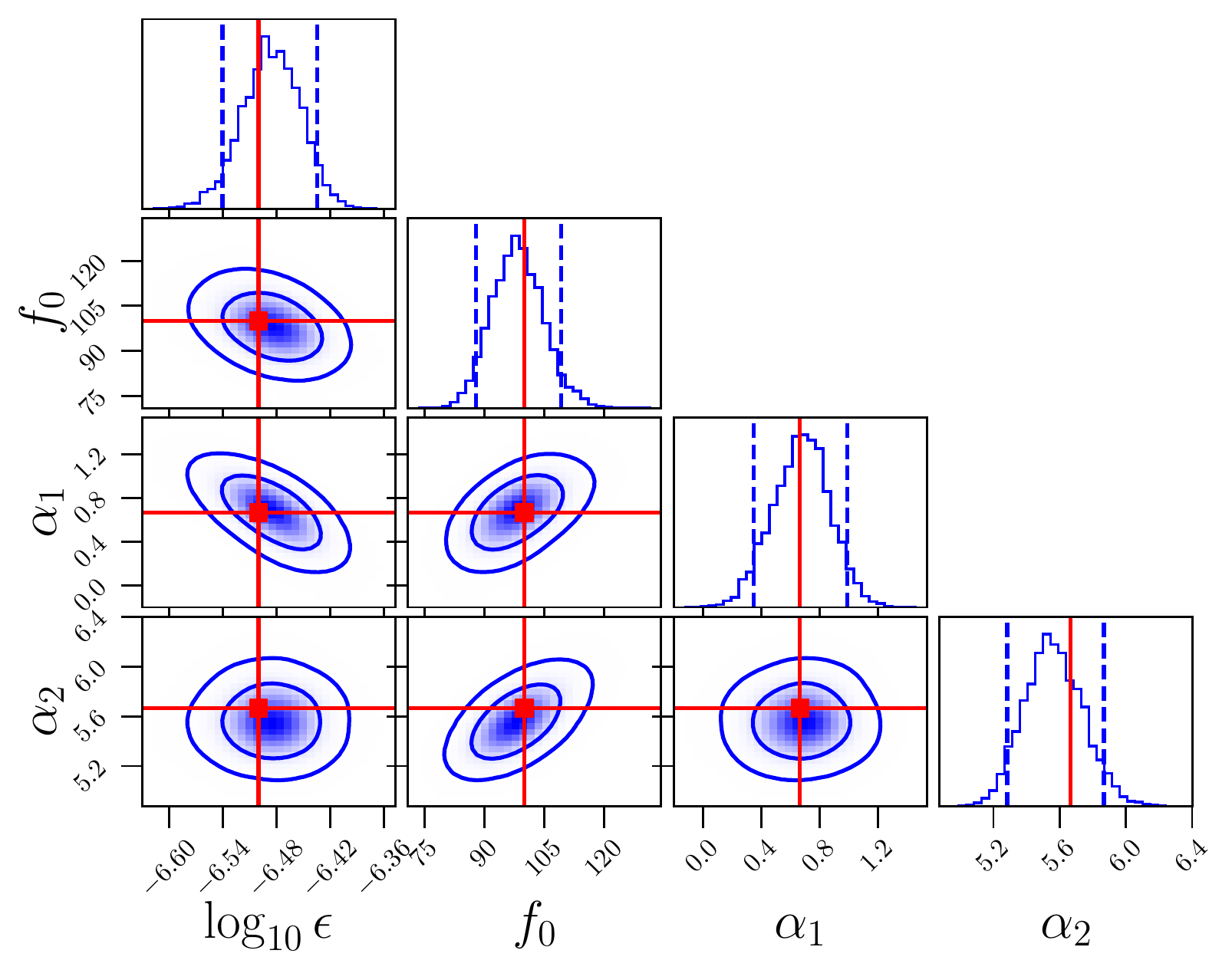}
      \caption{Example posterior of an injection recovery based on the \gls{bpl}
      model, with $\ln\mathcal{B}^{\rm BPL}_{\rm PL}=417$. The recovery is
      consistent with the injection, indicated by the red markers and lines.}
      \label{fig:ex_post_BPL}
    \end{figure}

    The first case of the model selection studies tests under which
    conditions our pipeline is able to distinguish a \gls{bpl} $\hf{}{}$
    from a \gls{pl}. While we follow the PL model shown in~\eqref{eq:H_PL},
    $\bar{H}_{\rm BPL}(f)$ is defined, prior to the normalization
    in~\eqref{eq:H_P_norm}, as
    \begin{equation}
      \bar{H}_{\rm BPL}(f)\propto\left\{
      \begin{array}{cc}
        (f/f_0)^{\alpha_1-3} & \text{if } f<f_0 \\
        (f/f_0)^{\alpha_2-3} & \text{if } f\geq f_0
      \end{array}\right..
    \end{equation}
    We model the \gls{bpl} case with four free parameters, i.e. three of them
    from $\bar{H}_\mathrm{BPL}(f)$ as well as the overall amplitude parameter,
    $\epsilon$. An example posterior is shown in \figref{fig:ex_post_BPL}. For
    this analysis, we inject the \gls{bpl} model with $\alpha_1=2/3$ and
    $f_0=100$ Hz fixed, hence exploring the $(\epsilon,\alpha_2)$ parameter
    space.  We adopt the prior distributions for each parameter as follows; for
    $\epsilon$, a log-uniform distribution from $10^{-12}$ to $10^{0}$, for
    $\alpha_1$ and $\alpha_2$, a Gaussian distribution with the mean of 0 and
    the standard deviation of 3.5, and for $f_0$, a uniform distribution in
    \SI{20}{\hertz} and \SI{500}{\hertz}.  The $\pbar_{\ell m}$ model is taken
    as the the sky distribution of the Galactic plane shown in
    \figref{fig:galactic_plane} with $\lmax=7$.

\begin{figure}[h]
      \centering
      \includegraphics[width=\columnwidth]{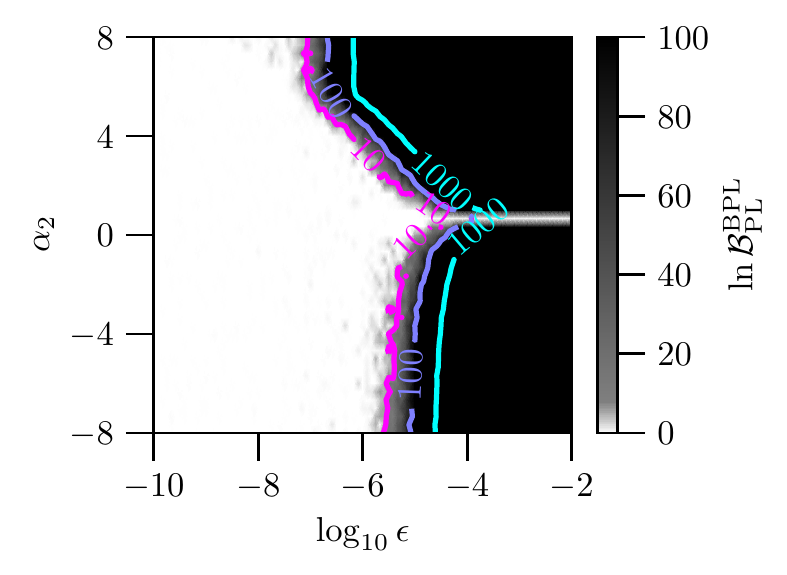}
      \caption{Heatmap showing the \gls{bf} for the \gls{bpl} vs \gls{pl}
      $\bar{H}(f)$ recovery models across grids of $\epsilon$ and $\alpha_2$
      values for injected \gls{bpl} $\bar{H}(f)$ with $f_0=100$ Hz,
      $\alpha_1=2/3$.}
      \label{fig:ms_bpl_heatmap}
    \end{figure}

    The heatmap in \figref{fig:ms_bpl_heatmap} suggests that higher values of
    $\epsilon$ result in the identification of the correct model with greater
    significance, transitioning around $\epsilon\sim 10^{-7}-10^{-6}$. The
    \gls{bf} tends to increase for higher values of $\alpha_2$. Also
    expectedly, the gap in $\alpha_2=2/3$ is observed because this is equal to
    the fixed value of $\alpha_1$, in which case the \gls{bpl} model becomes
    completely degenerated with a \gls{pl} of $\alpha=2/3$, leading to the equal
    odds between the two models $\mathcal{B}^{\rm BPL}_{\rm PL}\sim 1$.

  \subsection{\label{sec:model_selection-plm} $\pbar_{\ell m}$ model selection}
    Similar to the study summarized above, here we compare two different
    $\pbar_{\ell m}$ models, fixing $\bar{H}(f)$ to be a \gls{pl} model. We do
    not introduce any new free parameter in $\pbar_{\ell m}$ and explore only
    the ($\epsilon, \alpha$) parameter space.

\begin{figure}[h]
      \centering
      \includegraphics[width=\columnwidth]{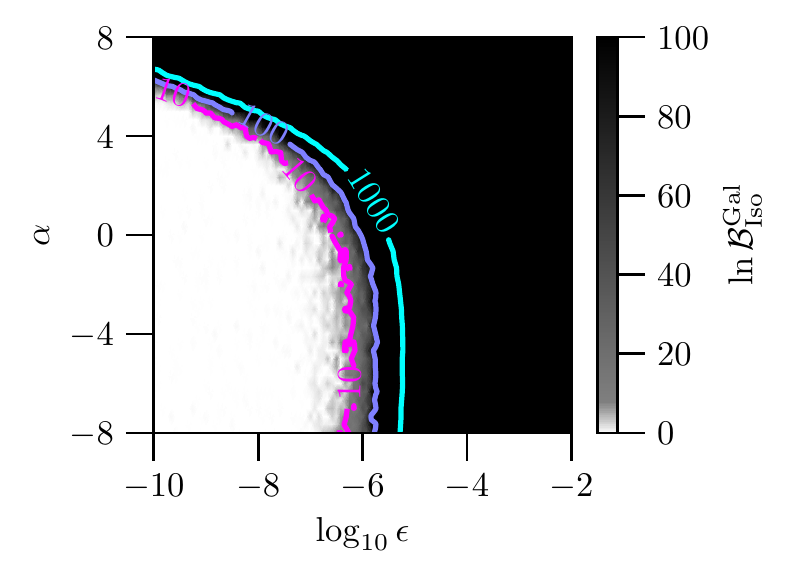}
      \caption{Heatmap showing the \gls{bf} for the Galactic plane vs
      isotropic $\pbar_{\ell m}$ recovery models across grids of $\epsilon$
      and $\alpha$ values for injected Galactic plane $\pbar_{\ell m}$ model.}
      \label{fig:ms_gal_heatmap}
    \end{figure}

    To demonstrate the capability of $\pbar_{\ell m}$ model selection, we
    inject a background signal of the Galactic plane $\pbar_{\ell m}$ with
    $\lmax=7$ and identify the parameter space where one can distinguish it from
    purely isotropic model, i.e.  $\pbar_{\ell m}=0$ if $\ell\neq 0$ or $m\neq
    0$. We note that given the normalization, an arbitrary $\pbar_{\ell m}$
    skymap with $\ell=0$ reduces to an isotropic model. We explore different
    injected values over the $(\epsilon,\alpha)$ parameter space, obtaining the
    heatmap shown in \figref{fig:ms_gal_heatmap}. The heatmap indicates that the
    analysis prefers the Galactic plane $\pbar_{\ell m}$ model at around
    $\epsilon=10^{-6}$ in the most conservative case, similar to the previous
    study.  This threshold decreases for high values of $\alpha$, which enhance
    the signal for high frequencies. Hence, for the darker part of the parameter
    space, the pipeline is capable of detecting the signature of higher-order
    spatial modes and distinguishing it from the isotropic background.
    
\section{\label{sec:lmax_opt} $\ell_{\mathrm{max}}$ optimization}
  As mentioned in \secref{sec:formalism-plm}, the likelihood in our formalism
  involves the angular scale cutoff, $\lmax$ as a hyperparameter to be tuned.
  The study in \appref{app:iso_bias} suggests the systematic bias in the
  parameter inference due to potential mismatch in $\lmax$ value. Also, it has been
  known that unreasonably high $\lmax$ causes overfitting to observed data, and
  thus several approaches have been taken to optimize the $\lmax$ value.  In the
  conventional \gls{shd} analysis [see e.g.,~\cite{o1_iso}], the $\lmax$ value is
  chosen in the consideration of the diffraction limit. However, this reasoning
  can be established only for a two-detector configuration, and also the $\lmax$
  choice depends on the shape of a signal spectrum, e.g., the power-law index of
  $\bar{H}(f)$. \citet{erik_angres} further investigated this in the case of a
  pointlike source with different signal amplitudes. The authors found that the
  larger $\lmax$ yields greater localization of the source, while the smaller
  $\lmax$ tends to recover larger \gls{snr}. This indicates the potential sweet
  spot of the $\lmax$ that compromises between the angular resolution and the
  signal detection, although the optimal choice potentially depends on the
  signal amplitude in the data. They also suggested that, for sufficiently loud
  signals, the $\lmax$ value can possibly surpass the one predicted from the
  diffraction limit argument. Yet, their conclusion cannot be directly applied to
  extended source models, which we consider in this work. Here, we describe a more
  generic and quantitative method to optimize the choice of $\lmax$ based on the
  Bayesian framework.

  \subsection{\label{sec:max_opt-method} Method}
    \citet{sph_methods} first discussed the approach to assess the
    validity of a chosen $\lmax$ by computing the Bayesian evidence given by
      \begin{align}
        p(\left\{C_{f t}\right\}|\lmax) = \int \boldsymbol{\diff}\pbar_{\ell m}\ p\left(\left\{C_{f t}\right\} \left|\pbar_{\ell m};\lmax\right.\right)p(\pbar_{\ell m}).
      \end{align}
    Although this approach can be applied, in principle, for arbitrary signal
    models or amplitudes, the number of marginalized parameters, which scales
    with $(\lmax+1)^2$, makes it practically unfeasible to evaluate the
    integration above with sufficient precision within a reasonable timescale.
    Instead, in the $\pbar_{\ell m}$-specific analysis, the Bayesian evidence is
    computed for a fixed $\pbar_{\ell m}$ model as well as $\lmax$ by
    marginalizing over only the model parameters, $\{\vec{\theta}\}$, which do
    not necessarily involve high dimensionality [see \eqref{eq:evidence}]. 
    
    Subsequently, we evaluate the odds ratio of a signal hypothesis
    ($\mathcal{S}$) over the noise hypothesis ($\mathcal{N}$), which reads as a
    function of $\lmax$
    \begin{align}
      \mathcal{O}^{\mathcal{S}}_{\mathcal{N}}(\lmax)&=\frac{p(\mathcal{S}, \lmax|\{C_{ft}\})}{p(\mathcal{N}|\{C_{ft}\})}\\
      &=\frac{p(\{C_{ft}\}|\mathcal{S}, \lmax)}{p(\{C_{ft}\}|\mathcal{N})}\frac{\pi(\mathcal{S}, \lmax)}{\pi(\mathcal{N})}.
    \end{align}
    Following the methodology in \secref{sec:model_selection-bf}, we assign
    equal prior odds to each signal and noise hypothesis and hence only
    consider the \gls{bf}
    \begin{align}
      \mathcal{B}^{\mathcal{S}}_{\mathcal{N}}(\lmax)=\frac{p(\{C_{ft}\}|\mathcal{S}, \lmax)}{p(\{C_{ft}\}|\mathcal{N})}
    \end{align}
    as the deciding factor.
    We note that the evidence for the noise hypothesis is given by the
    likelihood without any signal component subtracted, i.e. its exponent
    proportional to \eqref{eq:csdsq_whiten}. Since the Bayesian evidence
    quantifies the degree to which the real signal in the data can be described
    by the recovered signal model, we identify the $\lmax$ that maximizes
    $\mathcal{B}^{\mathcal{S}}_{\mathcal{N}}$ as its optimal choice.
  \subsection{\label{sec:max_opt-results} Results}
    We conduct a series of simulations using four kinds of the 1-year
    synthesized data with the projected O4 sensitivity as described in
    \secref{sec:stats-setup}. Three of them include a signal injection based on
    the \gls{pl} $\bar{H}(f)$ ($f_\mathrm{ref}=\SI{25}{\hertz}$ and $\alpha=4$)
    and the Galactic plane $\pbar_{\ell m}$ model shown in
    \figref{fig:galactic_plane} with the injected $\lmax=7$ and increasing
    amplitudes of $\epsilon=10^{-8}, 10^{-7}, 10^{-6}$.  The four datasets are
    analyzed by recovering the signal model consistent with the injection while
    varying $\lmax$ values from 0 to 10 for each dataset.  For the parameter
    estimation, we take $(\epsilon, \alpha)$ as free parameters to infer and set
    their priors as a log-uniform distribution from $10^{-13}$ to $10^{-5}$ and
    a Gaussian distribution with the mean of 0 and the standard deviation of
    3.5, respectively.

\begin{figure}[h]
      \centering
      \includegraphics[width=\columnwidth]{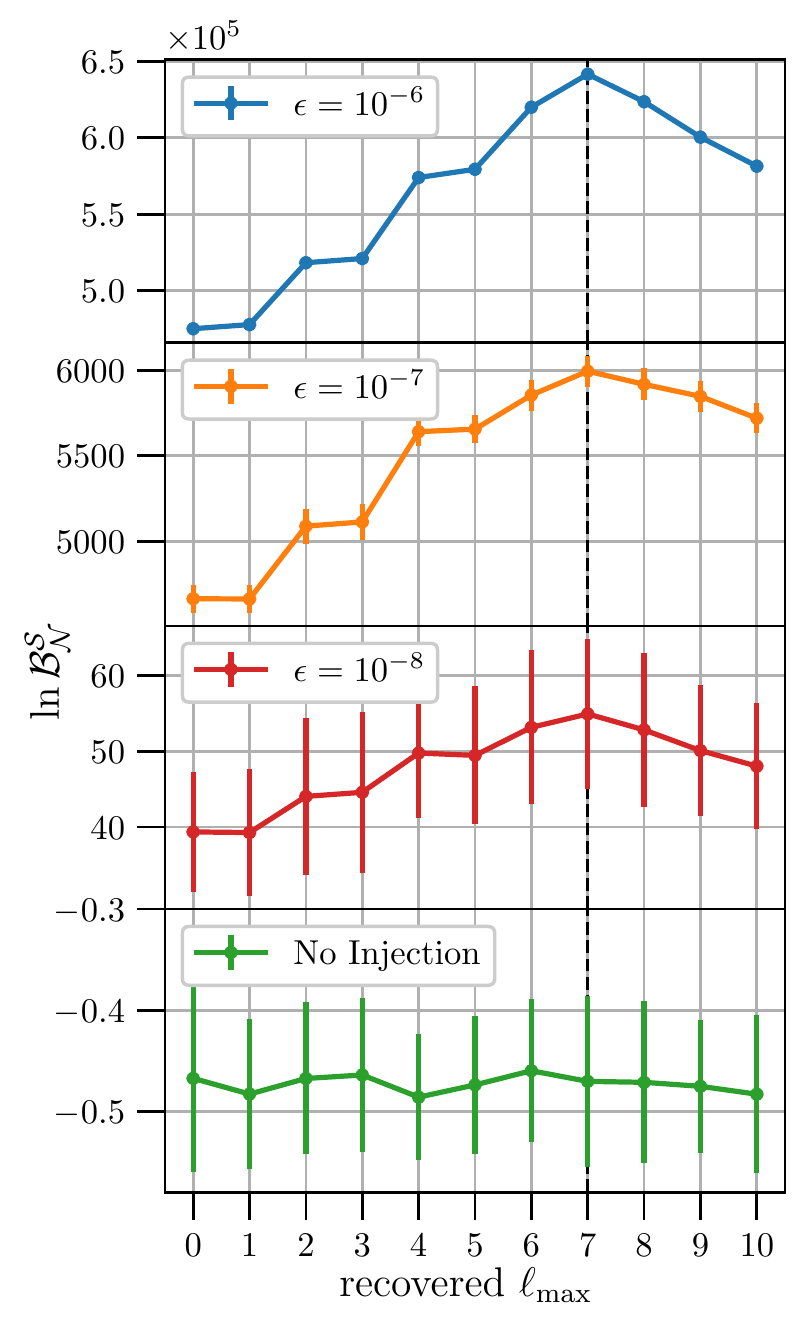}
      \caption{Recovered \glspl{bf} as a function of recovery $\lmax$. Different
      colors represent each magnitude of the background injection,
      $\epsilon=10^{-6}, 10^{-7}, 10^{-8}$ and no injection. The dashed black
      line at $\lmax=7$ is the $\lmax$ used for the injected signal model.}
      \label{fig:lmax_opt_compare}
      \vspace{-1in}
    \end{figure}

    For each dataset and the $\lmax$ value for recovery, the same analysis is repeated using
    10 different realizations of the noise data and the \gls{bf} is computed
    each time. \figureref{fig:lmax_opt_compare} shows its mean and 1-sigma error
    bar as a function of the recovered $\lmax$. We find that the \gls{bf} in log
    scale overall scales with roughly $\epsilon^2$ as expected from
    \eqref{eq:likelihood}, and its peak is located at the $\lmax$ value consistent
    with the injection ($\lmax=7$ indicated by the black dashed line) for all
    the three datasets.  The \gls{bf} for the noninjection dataset, on the
    other hand, does not indicate any significant detection and show a
    relatively flat structure across the recovered $\lmax$ values. Also, we note that
    the error bars tend to reduce relative to the mean for larger signal
    amplitudes. This observation implies that the optimal $\lmax$ is determined
    by the cutoff scale of the background signal present in data and that its
    significance increases with the signal amplitude.

\section{\label{sec:o3_zerolag} O3 MSP analysis}
  As an example of astrophysically motivated analyses beyond the toy models we
  described so far, we adopt a \gls{msp} $\pbar_{\ell m}$ model and search for
  the signal model using the publicly available
  data~\cite{foldeddata_release} from \gls{o3} of the \gls{lvk}.  We apply
  the same preprocessing and gating to the \gls{o3} timeseries data collected
  from the two \gls{ligo} and Virgo detectors as described
  in~\cite{o3_directional}, resulting in the \gls{csd} with different livetime
  for each detector pair (i.e. 169~days for HL, 146~days for HV and 153~days for
  LV baseline respectively). Subsequently, the \gls{csd} is folded to one
  sidereal day using the method in~\cite{folding}, and the static data products
  are computed by the \texttt{PyStoch} pipeline~\cite{pystoch, pystoch_sph} and
  stored following the procedure similar to \secref{sec:data-noise}.
  \begin{figure}[h]
    \centering
    \includegraphics[width=.9\columnwidth]{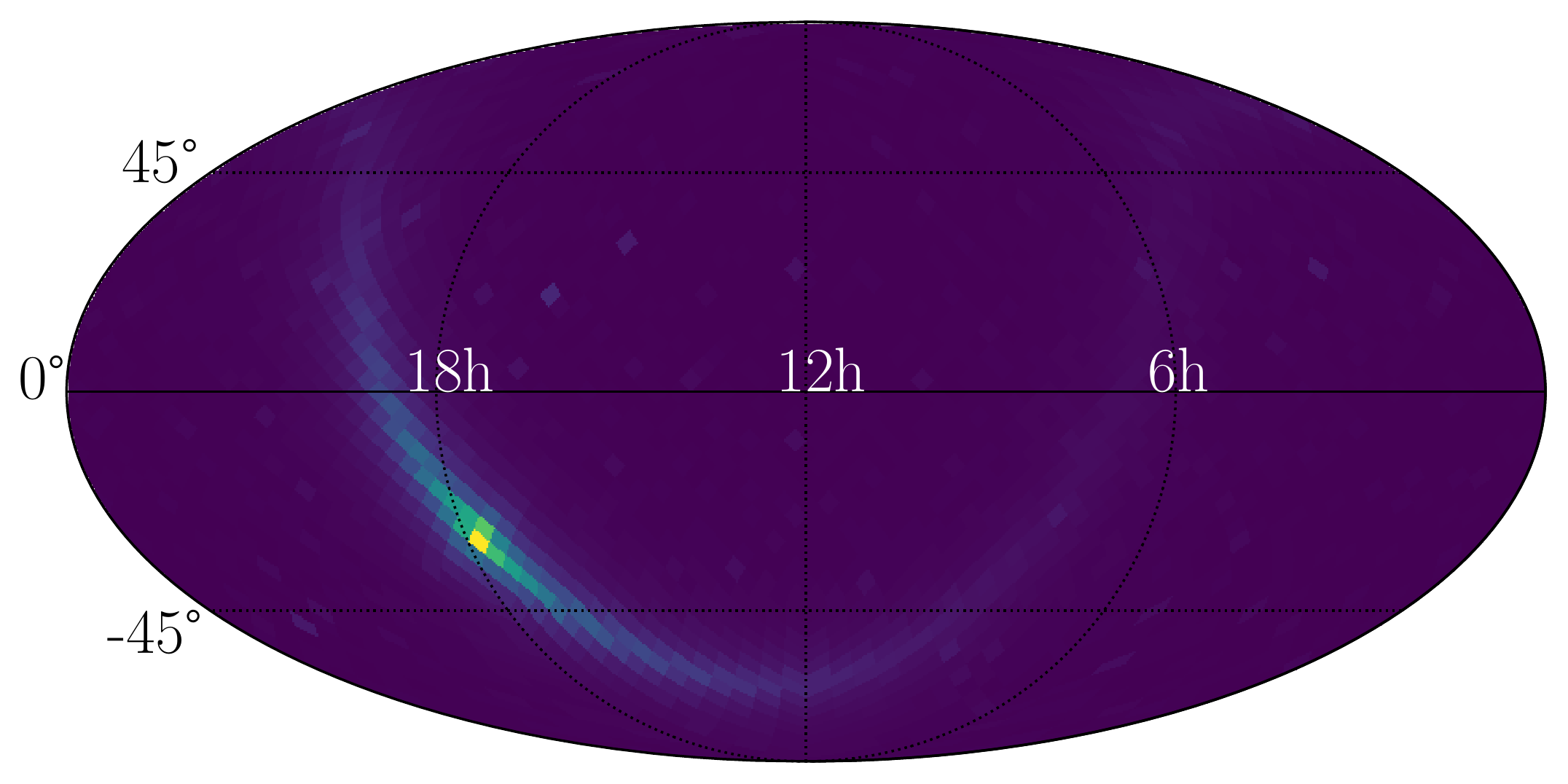}
    \caption{Distribution of the \gls{msp} population visualized by Mollweide projection with \texttt{HEALPix}~\cite{HEALPix}
    pixelization of $n_\mathrm{side}=16$. The brighter color represents larger
    energy density.}
    \label{fig:galactic_plane_msp}
  \end{figure}
  
  Previously, similar searches for the \glspl{msp} were conducted based on the
  isotropic~\cite{10.1093/mnras/stac984} or targeted~\cite{deepali_targeted}
  analysis. The both searches provide the upper limit on averaged ellipticity of
  the \gls{msp} population and the number of \glspl{msp} in the observed
  frequency band.  Regarding the \gls{msp} population in our analysis, we adopt
  the $\pbar_{\ell m}$ model developed in~\cite{deepali_targeted} using the
  Gaussian density profile for the pulsar radii, as shown in
  \figref{fig:galactic_plane_msp}. Note that the model developed for the sky
  distribution is in the pixel basis ({\tt HEALPix}~\cite{HEALPix} grid) and can
  be converted to the spherical harmonic basis. The same spectral model follows
  $\bar{H}(f)=f^4p(f|\mu, \sigma)$ where $p(f|\mu, \sigma)$ is a probability
  density function of the log-Gaussian form with the mean of $\mu$ and the
  standard deviation of $\sigma$. Unlike~\cite{deepali_targeted}, we set
  $\epsilon$ and $\mu$ as free parameters to sample, which obey the priors of a
  log-uniform distribution from $10^{-15}$ to $10^{-5}$ and a Gaussian
  distribution with the mean of 6.1 and the standard deviation of 0.2
  respectively, while $\sigma$ is fixed to be 0.58. These choices of $\mu$ and
  $\sigma$ are motivated by the best fit values found in \cite{Lorimer_2015}. 
  
\begin{figure}[h]
    \centering
    \includegraphics[width=.9\columnwidth]{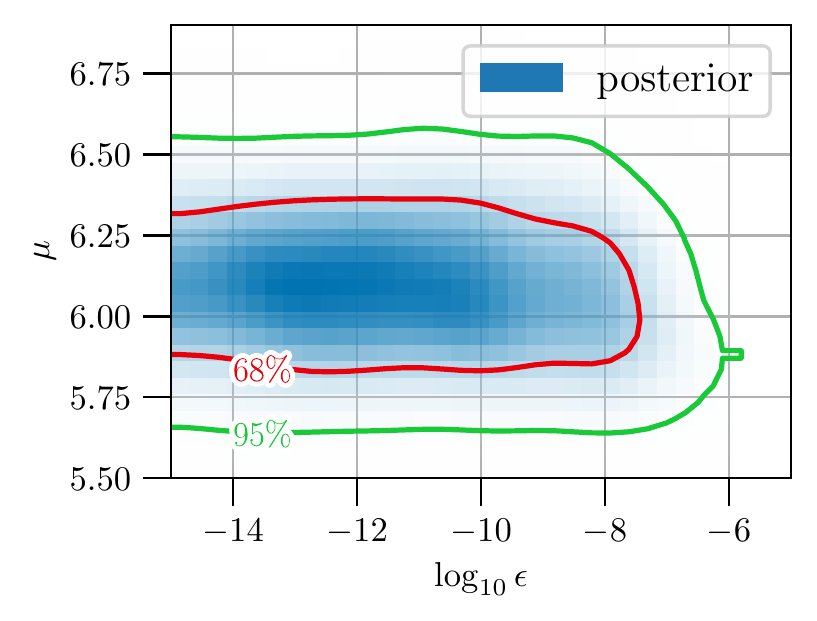}
    \caption{Posterior distribution of $(\epsilon, \mu)$ parameters for the
    \gls{msp} population with $\lmax=5$ using the HLV data.}
    \label{fig:msp_lmax5_posterior}
  \end{figure}

  We analyze the O3 data with different $\lmax$ values from 0 to 5, evaluating
  the likelihood without any injection, \eqref{eq:likelihood2}, which provides
  the \gls{bf} and posterior results. The computed \gls{bf} values do not have
  any strong dependence on $\lmax$ values, fluctuating around $-0.2$.
  Therefore, we do not identify any evidence of the \gls{sgwb} from the
  \gls{msp} population.  \figureref{fig:msp_lmax5_posterior} shows the posterior
  result from the analysis with $\lmax=5$ using the HLV data, which indicates
  that the posterior of $\mu$ is similar to its prior and hence that we do not
  obtain its meaningful constraints. We find that the 95\% upper limit of the
  amplitude $\epsilon\leq 2.7\times 10^{-8}$ with the HLV data, being in good
  agreement with~\cite{deepali_targeted}. We also note that the posterior
  results given by other $\lmax$ values produce a similar structure and the
  upper limits on $\epsilon$.  \footnote{The $95\%$ confidence upper limit from
  \cite{deepali_targeted} is $\epsilon\leq 6.7\times 10^{-8}$ assuming a
  log-uniform prior with the same range as mentioned in our analysis.  The
  difference may arise from the sky resolution used in the analysis and
  variation of $\mu$.} 
  
\section{\label{sec:future} future prospects}
Given the current sensitivity of ground-based \gls{gw} detectors, it is
challenging to detect any kind of anisotropic \gls{sgwb} at this point. Rather, it would be of
our greater interest to assess the future prospect of detectability or
constraints on model parameters using projected sensitivity of current and
next-generation detectors. Here, we present the results of simulated analyses
using three different baseline configurations as follows:
\begin{itemize}
  \item the two \gls{ligo}, Virgo and KAGRA detectors with their design sensitivities (HLVK design)
  \item the two \gls{ligo}, Virgo, KAGRA and LIGO-India detectors with the A+ sensitivities (HLVKI A+)
  \item the two \gls{ligo}, Virgo, KAGRA and LIGO-India detectors with the A+
  sensitivities as well as Cosmic Explorer (HLVKI+CE)
\end{itemize}

\begin{figure}[h]
  \centering
  \includegraphics[width=\columnwidth]{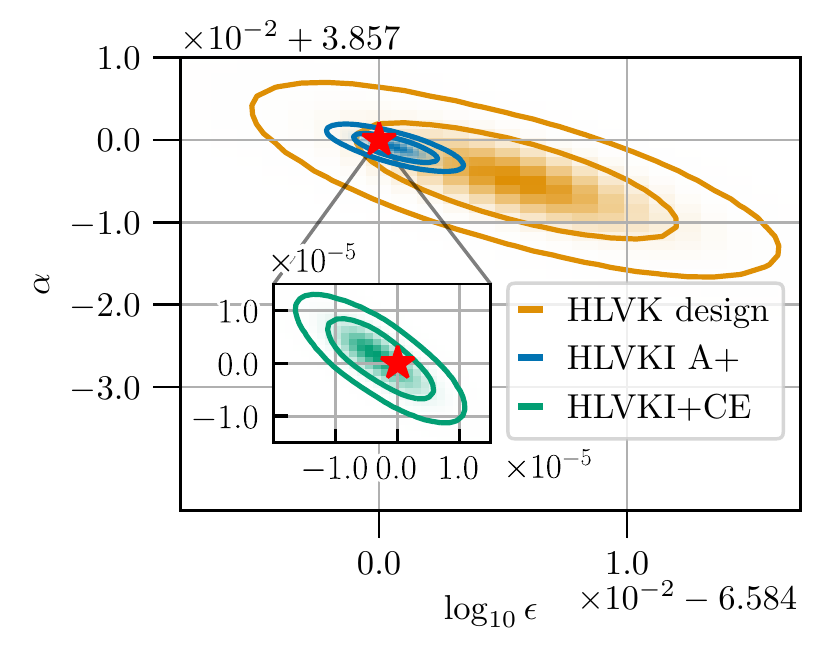}
  \caption{Posterior distribution for the same detected injection as
  \figref{fig:ex_post_detected}. The inner and outer contour for each
  distribution represent 1$\sigma~(68\%)$ and 2$\sigma~(95\%)$ credible
  region.}
  \label{fig:future_HLVK_HLVKI_HLVKICE_inj}
\end{figure}

For each 1 sidereal-day dataset, we conduct two analyses with and without a background signal
injected. The injected signal model follows the \gls{pl} $\bar{H}(f)$ and
the Galactic plane $\pbar_{\ell m}$ model with the same injected parameters
$(\epsilon, \alpha)=(2.608\times10^{-6}, 3.857)$ as the one in
\secref{sec:stats-recovery} and $\lmax=7$. This injected
signal is recovered with the consistent signal model using the same prior
distribution as the injection recovery in \figref{fig:ex_post_detected}, and the
2D posterior result for each dataset is shown in
\figref{fig:future_HLVK_HLVKI_HLVKICE_inj}, where the injected values are
indicated by the red star. The inner and outer contours represent the $65\%$ and
$95\%$ confidence region, respectively.

We find the \gls{bf} of these signal recoveries to be roughly
$10^{5},10^{7},10^{12}$ for the injection in the ``HLVK design'',``HLVKI
A+'',``HLVKI+CE'' datasets, respectively. Regarding the uncertainty in parameter
inference, the ``HLVK design'' network improves it by a factor of 2 compared to
that shown in \figref{fig:ex_post_detected}, where the network configuration
contains only the two LIGO detectors with the same design sensitivity.
Furthermore, the uncertainties for ``HLVKI A+'' (``HLVKI+CE'') network decrease
by another one (three) order(s) of magnitudes. Nevertheless, we should note that
these future configurations would be so sensitive that the energy spectrum of a
background signal itself might act as another source of noise background. Since
the current formalism does not account for the effect of this additional noise,
we regard the results shown in \figref{fig:future_HLVK_HLVKI_HLVKICE_inj} as
highly optimistic and leave more thorough analysis without the weak-signal
assumption as future work.

\begin{figure}[h]
  \centering
  \includegraphics[width=\columnwidth]{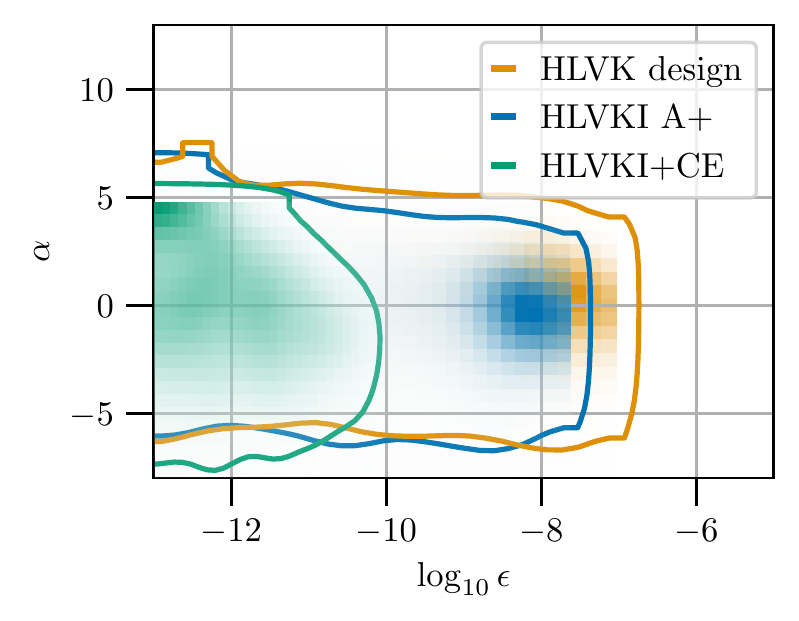}
  \caption{Posterior distribution for dataset using each future detector network
  without any injection. The contours for each distribution represent
  2$\sigma~(95\%)$ credible region for parameter constraints.}
  \label{fig:future_HLVK_HLVKI_HLVKICE_noiseonly}
\end{figure}

On the other hand, the results for the noninjection analysis are presented in
\figref{fig:future_HLVK_HLVKI_HLVKICE_noiseonly}.  For each dataset, we use the
same prior distribution as the signal recovery in
\figref{fig:ex_post_nondetected} and draw a contour that represents the $95\%$
confidence region. While all the datasets yield similar constraints for
$\alpha$, the upper limit of $\epsilon$ becomes more stringent by a half (three)
order(s) of magnitudes for ``HLVKI A+'' (``HLVKI+CE'') compared to
\figref{fig:ex_post_nondetected}. This demonstrates how significantly the
constraints on the background amplitudes would improve for next generations of
\gls{gw} detector networks.

\section{\label{sec:conclusion} Conclusion}
  In this paper, we introduced a parameter estimation formalism targeted for an
  anisotropic \gls{sgwb} from extended sources. Following the Bayesian
  framework, this formalism provides the posterior distribution of model
  parameters for a given anisotropy model $\pbar_{\ell m}$, which can be seen as
  an extension from the Bayesian parameter estimation for isotropic \gls{sgwb}
  with higher spatial modes in a signal model. We also developed data
  simulation and signal injection tools on the spherical-harmonics bases, which
  allow us to conduct injection studies to assess statistical consistency of
  posterior results. The study we conducted using the Galactic plane model showed
  validity of the posterior results given by our analysis [see
  \figref{fig:pp_plot}].
  
  Also, this formalism brings additional novelties such as a model selection
  between different $\pbar_{\ell m}$ or $\bar{H}(f)$ models and a systematic
  method to optimize $\lmax$ value for given $\pbar_{\ell m}$, both of which
  make use of the comparison of the \gls{bf}. In particular, we explored the
  parameter region where our pipeline can distinguish two different signal
  models in various cases, e.g., \gls{pl} and \gls{bpl} $\bar{H}(f)$ models, or
  isotropic and the Galactic plane $\pbar_{\ell m}$ models. We noted that the
  structure of the distinguishable parameter region was consistent with our
  expectation. [see Figs.~(\ref{fig:ms_bpl_heatmap},\ref{fig:ms_gal_heatmap})]
  Regarding the $\lmax$ optimization, we found that, throughout injected signal
  amplitudes that are loud enough, the recovery $\lmax$ maximizes the \gls{bf}
  when it was matched with the $\lmax$ value used for the signal injection. [see \figref{fig:lmax_opt_compare}]

  Finally, we conducted an analysis to search for the \gls{msp} population using
  the real data from the \gls{lvk}'s \gls{o3}. We didn't have any strong
  evidence of such a \gls{sgwb} signal in the data and placed the constraints on
  the amplitude $\epsilon\leq 2.7\times10^{-8}$ (95\% upper limit).  We also
  simulated analyses using the projected sensitivities of future \gls{gw}
  detector networks. The analyses demonstrated that the upgrade to the A+
  sensitivity or the addition of CE improves the precision of parameter
  estimation by one or three orders of magnitudes, respectively. We also
  assessed the constraints potentially placed on the same parameters and obtain
  similar degrees of the improvements. This indicates promising insights into
  the search for anisotropic \glspl{sgwb} in the future observations.

\begin{acknowledgments}
  We would like to thank Jishnu Suresh and Vuk Mandic for their fruitful
  discussion and feedback.  L.T is supported by the National Science Foundation
  through OAC-2103662 and PHY-2011865. S.J. is supported by grants
  PRE2019-088741 funded by MCIN/AEI/10.13039/501100011033 and FSE,
  PGC2018-094773-B-C32 [MCIN-AEI-FEDER] and CEX2020-001007-S [MCIN]. The authors
  are also grateful for computational resources provided by the LIGO Laboratory
  and supported by National Science Foundation Grants PHY-0757058 and
  PHY-0823459. This material is based upon work supported by NSF's LIGO
  Laboratory which is a major facility fully funded by the National Science
  Foundation.
\end{acknowledgments}

\appendix

\section{\label{app:iso_bias} POTENTIAL BIAS IN ISOTROPIC ANALYSIS}
In the injection studies shown in \secref{sec:stats}, we adopt the same $\lmax$
for both the injection and its recovery, and verify the statistical consistency
in the parameter inference.  In reality, however, we do not know the spatial
scale of a background signal in nature \textit{a priori}, and it is possible that
one does not recover the signal with a consistent $\lmax$ value. Especially,
isotropic \gls{sgwb} searches (i.e. $\lmax=0$ for recovery) always ignore any
higher spherical-harmonics mode and hence, for future configurations with
greater sensitivities, it is crucial to investigate the potential impact of the
$\lmax$ inconsistency to the parameter inference. Here, we describe the results
of an injection study similar to what is shown in \secref{sec:stats-pp_plot}. 

\begin{figure}[h]
  \centering
  \includegraphics[width=\columnwidth]{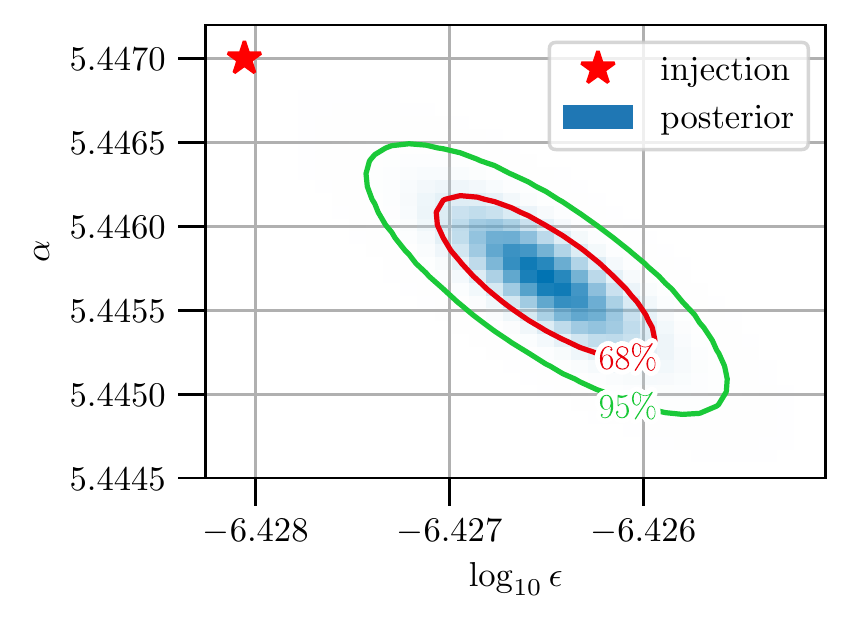}
    \caption{Posterior distribution for one of the detected injections that
    indicate a strong bias in the signal recovery. The injected values are
    $(\log_{10}\epsilon, \alpha)=(-6.428,5.447)$ and the \gls{bf} is
    $\sim5.6\times 10^{7}$.}
  \label{fig:iso_bias_post}
\end{figure}
We use the same dataset with injections produced from the same signal model and
parameter values of ($\epsilon, \alpha$) as used for \secref{sec:stats-pp_plot},
setting $\lmax=3$. Each of these injections is recovered with the consistent
signal model as well as the same prior distribution except $\lmax=0$ to simulate
an isotropic \gls{sgwb} analysis. \figureref{fig:iso_bias_post} exemplifies a
typical bias of the parameter inference seen in one of the 2D posterior results.
Given the relatively large \gls{bf} ($\sim 6\times 10^{7}$) found for the injection
shown in \figref{fig:iso_bias_post}, we note that this bias tends to be more
noticeable for louder injections.

\begin{figure}[h]
  \centering
  \includegraphics[width=\columnwidth]{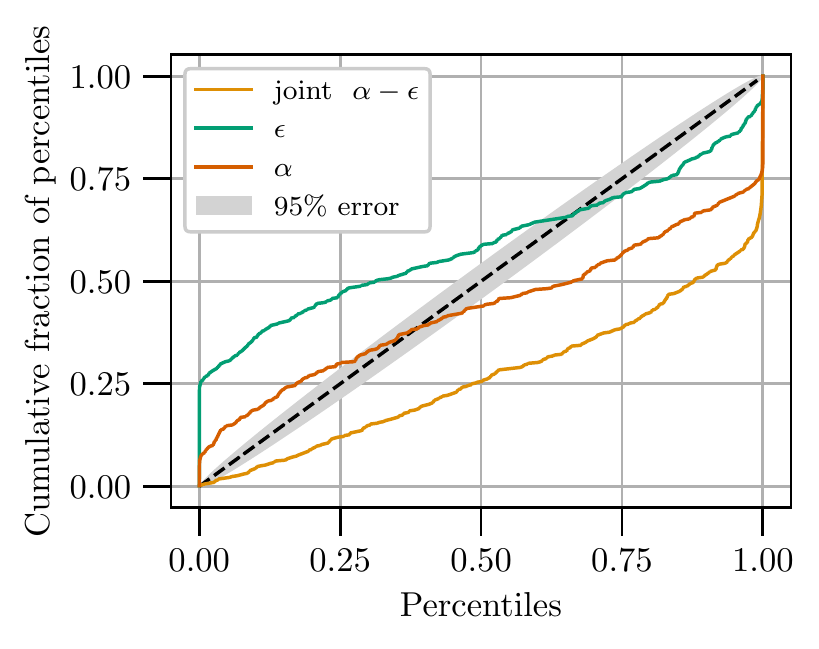}
  \caption{Probability-probability plot for 500 of the injection recoveries
  using the recovery $\lmax=0$.  Different colors indicate each method of
  computing the percentile by either 1D marginalized or 2D joint distribution.
  The gray region is the 95\% confidence region expected from the Poisson
  fluctuation.}
  \label{fig:iso_bias-pp_plot}
  \end{figure}
In order to assess a statistical picture, a P-P plot for these injection
recoveries is produced and shown in \figref{fig:iso_bias-pp_plot}. Apart from
the percentile computed for the 2D $\alpha - \epsilon$ posterior as explained in
\secref{sec:stats-pp_plot}, we also plot percentiles evaluated for each of the
1D marginalized posteriors. Unlike the $\lmax=3$ case in \figref{fig:pp_plot},
all three curves clearly deviate from the $95\%$ error region. Therefore,
this result suggests that the parameter inference is systematically biased when
the recovery $\lmax$ is not consistent with the one characterizing a real
signal, which will be particularly impactful in the conventional isotropic
analysis in the future.  Aiming to avoid this bias, we discuss a quantitative
approach to optimize the recovery $\lmax$ in \secref{sec:lmax_opt}.


\bibliography{references}

\end{document}